# One-dimensional vertical dust strings in a glass box

Jie Kong, Truell W. Hyde, Lorin Matthews, Ke Qiao, Zhuanhao Zhang, Angela Douglass

Center for Astrophysics, Space Physics, and Engineering Research (CASPER)

Baylor University

Waco, Texas 76798-7310

E-mail: J_Kong@baylor.edu; Truell_Hyde@baylor.edu

Abstract

The oscillation spectrum of a one-dimensional vertical dust string formed inside a glass box on top of the lower electrode in a GEC reference cell was studied. A mechanism for creating a single vertical dust string is described. It is shown that the oscillation amplitudes, resonance frequencies, damping coefficients, and oscillation phases of the dust particles separate into two distinct groups. One group exhibits low damping coefficients, increasing amplitudes and decreasing resonance frequencies for dust particles closer to the lower electrode. The other group shows high damping coefficients but anomalous resonance frequencies and amplitudes. At low oscillation frequencies, the two groups are also separated by a $\pi$-phase difference. One possible cause for the difference in behavior between the two groups is discussed.

I. INTRODUCTION

One-dimensional systems are of great interest in both theoretical physics and applied engineering research [1 – 8]. In engineering research, much of nano-scale technology (for example the production of quantum wires and/or carbon nanotubes), can also be related to a one-dimensional system. Dusty plasmas provide a controllable environment that can easily be used to simulate such systems. For example, low dimensional transport and transition phenomena (e.g., thermal conductivity, wave propagation, and phase transitions) are all easily produced and controlled within a dusty plasma environment due to its rapid response time as compared to a colloidal system and the fact that a dusty plasma can be easily imaged, unlike atomic systems.

Theoretically, one-dimensional dust strings have been investigated by many authors [9 – 13]. However, to date, controllable one-dimensional dust strings have not been extensively examined experimentally due to the difficulty of forming them in a complex plasma, especially in the vertical direction [14]. In this paper, a technique is provided for generating a one-dimensional vertical dust string at the center of a glass box placed on the lower electrode of a GEC rf reference cell [15]. Once in place, this one-dimensional string is levitated in the plasma sheath above the lower electrode and its oscillation spectra examined through modulation of an external DC bias applied to the lower electrode.

## II. EXPERIMENT AND RESULTS

*Overview of the experiment*

The experiment described in this work was conducted within a modified Gaseous Electronics Conference (GEC) rf reference cell located in the CASPER lab at Baylor University [16]. The CASPER GEC rf cell produces a capacitively coupled rf discharge formed between two parallel-plate electrodes in Argon gas. The lower electrode is 8 cm in diameter, air-cooled and separated from the upper electrode by 3 cm, and is powered by an rf signal generator. The upper electrode is grounded, as is the chamber. Melamine formaldehyde dust particles, $8.89 \pm 0.09$ μm in diameter, are introduced into the plasma via shakers mounted above the hollow upper electrode.

As particles fall through the plasma, they acquire a negative charge, attaining equilibrium within the sheath region above the lower electrode. For this experiment, horizontal confinement of the dust is produced employing an open-ended square glass box of height H = 12 mm, side length L = 10.5 mm, and wall thickness of 2 mm (Figure 1). These glass box dimensions are in general smaller than those used in most other recent experiments [17 – 19].

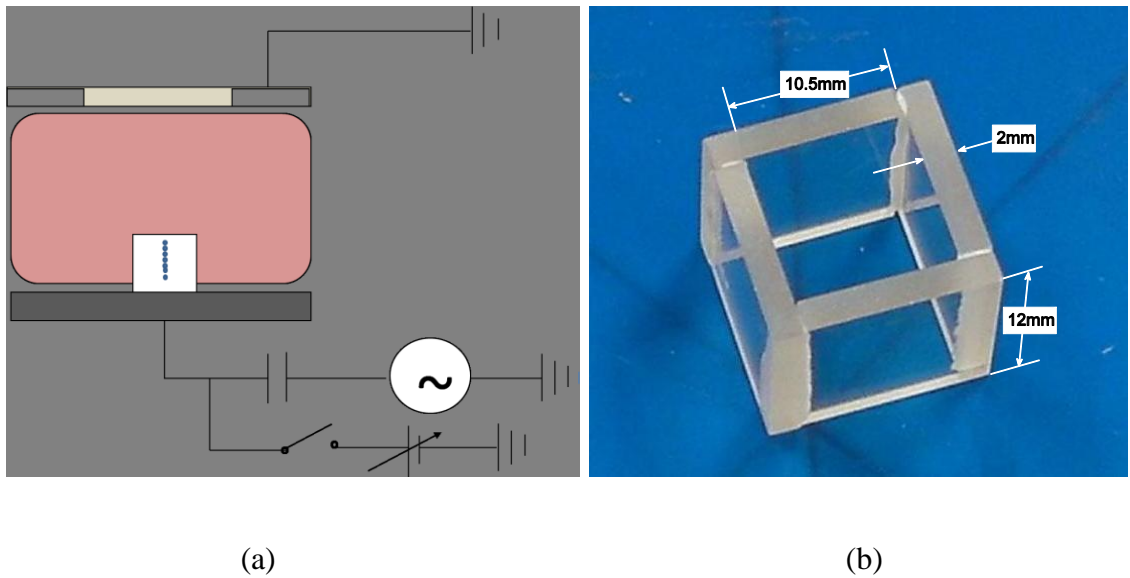

(a) (b)

Figure 1. (Color online) (a) Schematic of the experimental setup. (b) The open-ended square glass box used in the experiment described in the text.

To form a single vertical dust string, the rf power was initially established at 5 W, at a gas pressure of 150 mTorr. Under the parameters described, dust particles introduced into the system first formed a turbulent cloud close to the upper edge of the box. Slowly lowering the rf power to approximately 3.6 W caused this turbulent dust cloud to pass through a filamentary state [20] before forming a single long vertical string located at the center of the glass box (Figure 2). The total number of particles within this one-dimensional string was always between 10 and 20. Upon examination, it was determined that the transition from a turbulent state to a 1D vertical string was reversible: returning the rf power to its original value caused the 1D string to transition back through the filamentary state to the turbulent state. Finally, continually lowering the rf power caused the dust particles to fall to the lower electrode, starting with particles closest to the electrode.

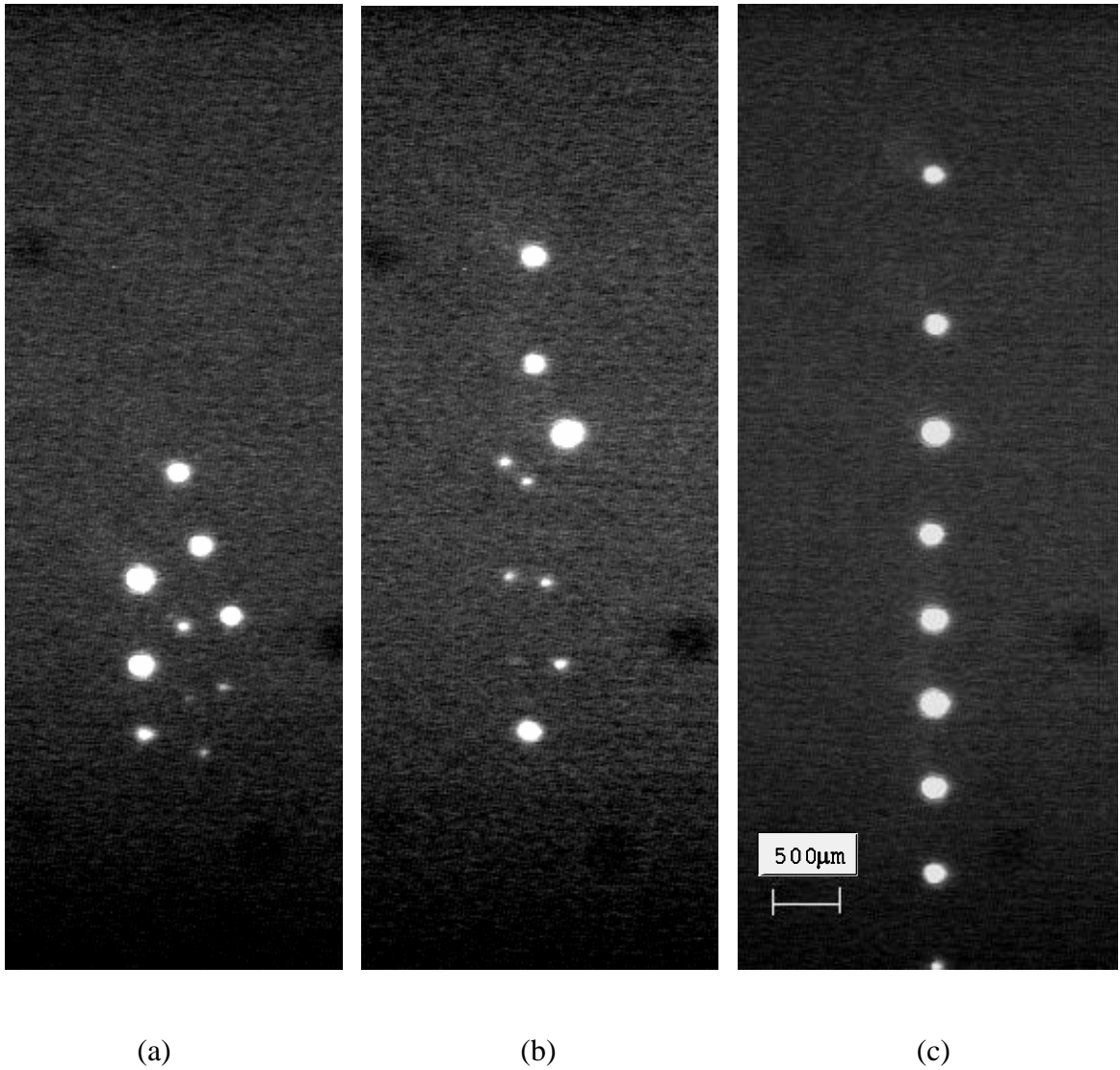

(a) (b) (c)

Figure 2. Transition from filamentary state to a single vertical string by varying the rf power. In the figure above, the rf power decreases from (a) to (c).

*Interaction of the glass box with the one-dimensional dust string*

As mentioned above, lowering the rf power causes a dust particle structural transition from 3D to 1D but only when the particles are confined in a glass box, indicating that the horizontal confinement increases as the rf power decreases. This can best be explained by the fact that lowering the rf power decreases the ionization rate, increasing the screening length. Since in this case horizontal confinement is due to the surface charge on the glass box [21], the Yukawa force exerted on the dust particles by each of the glass panes increases as the screening length increases, increasing the overall horizontal confinement. Due to symmetry, this results in the overall horizontal confinement canceling at the center of the box, creating an ideal environment for one-dimensional vertical string formation.

*Oscillation of one-dimensional dust strings*

Using the technique described above, one-dimensional vertical strings consisting of eight, six, four, and two particles were assembled within the box. Once stable, they were perturbed employing a low frequency (1 – 20 Hz) sine wave produced by an Agilent 33120A waveform generator which provided a modulated DC bias to the lower electrode. The resulting resonance spectra were examined and then compared to that found for a single, isolated particle. Strings with eight or fewer particles were used in this experiment in order to maximize overall system stability; at the same time, input modulation amplitudes were attenuated in order to avoid nonlinear dust particle oscillations. Particles were illuminated using a vertically fanned laser sheet and system data were captured employing a side-mounted CCD camera running at 60 frames per second.

*Experimental results*

Throughout this paper, particles are numbered P1 through P8, with P1 being the particle farthest away from the lower electrode (i.e., the particle at the top of the string). Representative oscillation spectra for a string composed of eight particles are shown in Figure 3(a). A series of resonance peaks for each particle can be seen, with the minimum amplitude peak occurring for P2. For comparison purposes the oscillation spectrum for a single particle along with a theoretical curve fit are shown in Figure 3(b).

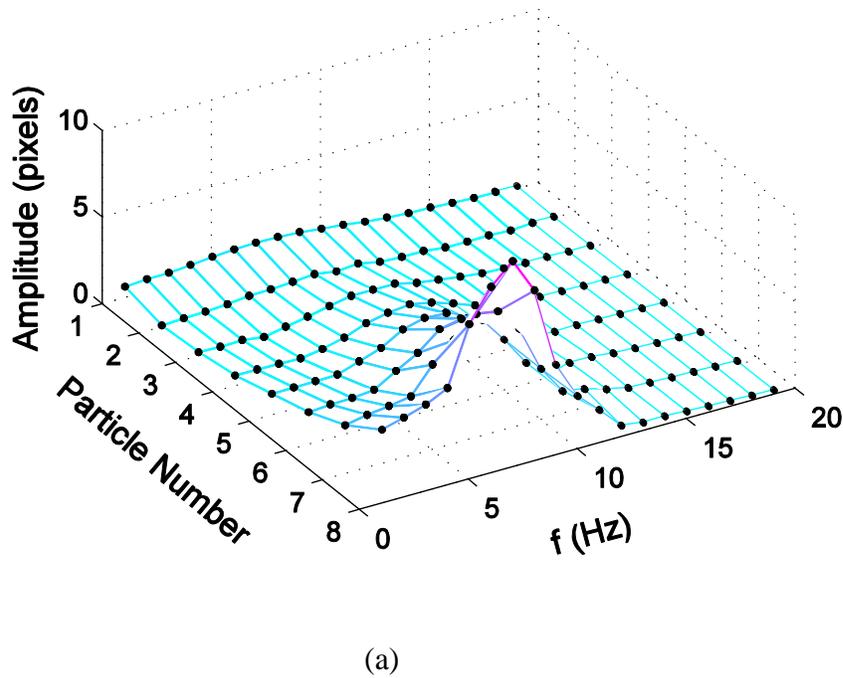

(a)

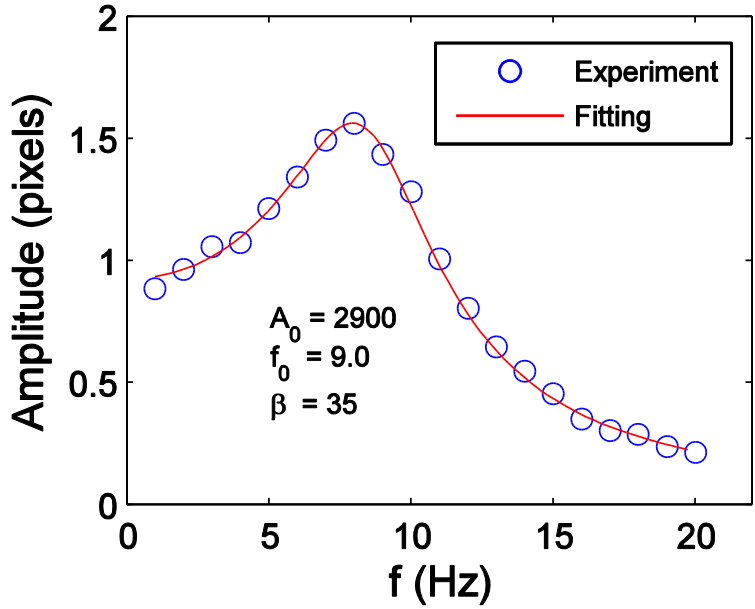

(b)

Figure 3. (Color online) (a) Representative oscillation spectra for a vertical dust string consisting of eight particles. Particles are numbered with larger numbers representing dust particles in the string positioned closest to the lower electrode. (b) Oscillation spectrum of a single particle with a theoretical fit used to calculate the amplitude $A_0$, resonant frequency $f_0$, and damping coefficient $\beta$ (in unit $\beta/2\pi$ = Hz).

The theoretical fit shown in Figure 3(b) was derived employing the response amplitude fitting,

$$A_i = \frac{A_{0i}}{\sqrt{\left(\omega_{0i}^2 - \omega^2\right)^2 + \beta^2 \omega^2}} \qquad (1)$$

In Eq. 1, $\beta$ is the damping coefficient (in unit $\beta/2\pi$ = Hz). The response amplitudes, resonance frequencies, and damping coefficients to be determined for each particle in the string are shown in Fig 4 for the representative case under consideration.

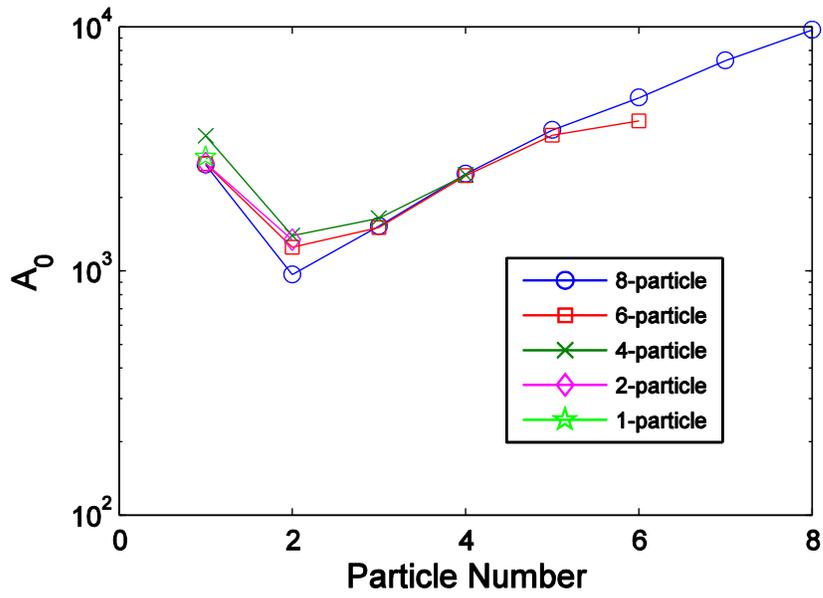

(a)

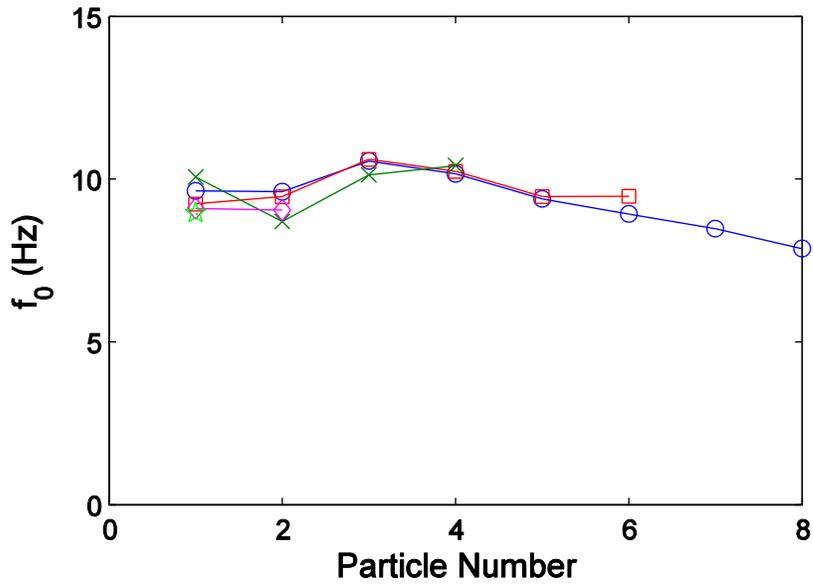

(b)

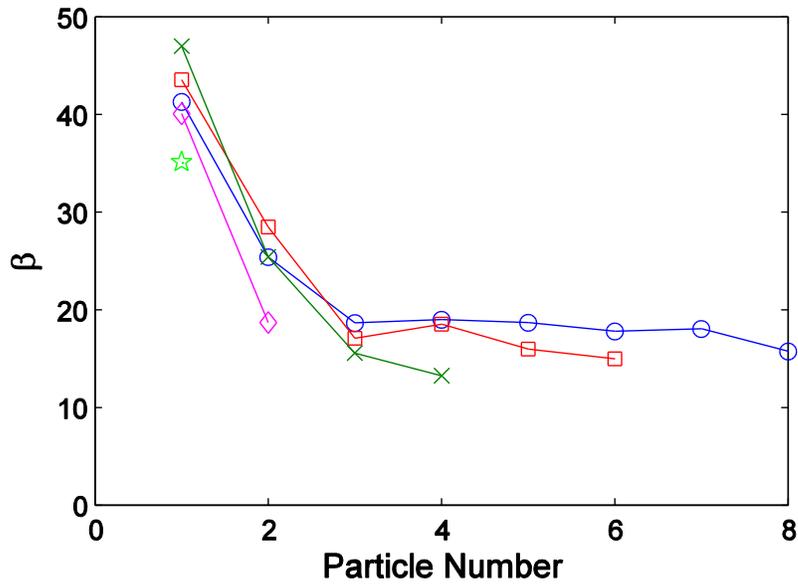

(c)

Figure 4. (Color online) Comparison of experimentally determined (a) response peak amplitudes, (b) resonance frequencies, and (c) damping coefficients for different sized dust strings. A particle-by-particle comparison shows that all three parameters remain approximately constant for a given position within the string. It is important to mention that $\beta/2\pi \approx 3 Hz$ for particles P3 – p8, which is small compared to the resonance frequency (~ 9Hz). However, damping for P1 and P2 are around 7 Hz, close to but still less than critical damping.

The oscillation phases for each particle within different sized strings are shown in Fig 5 where phases are measured with respect to the phase of P1.

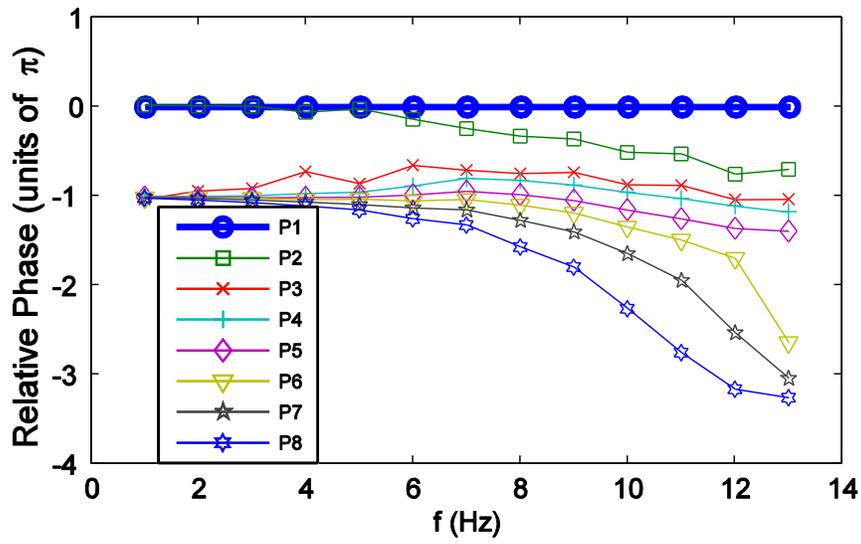

(a)

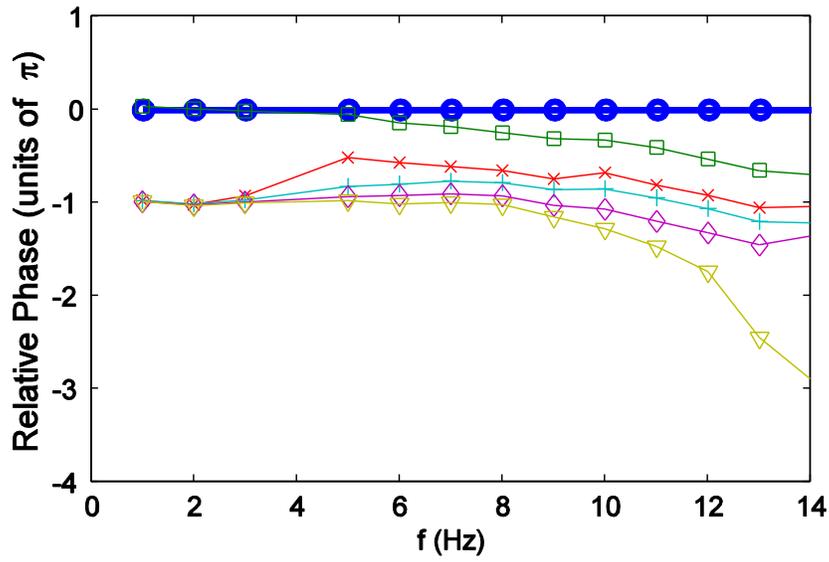

(b)

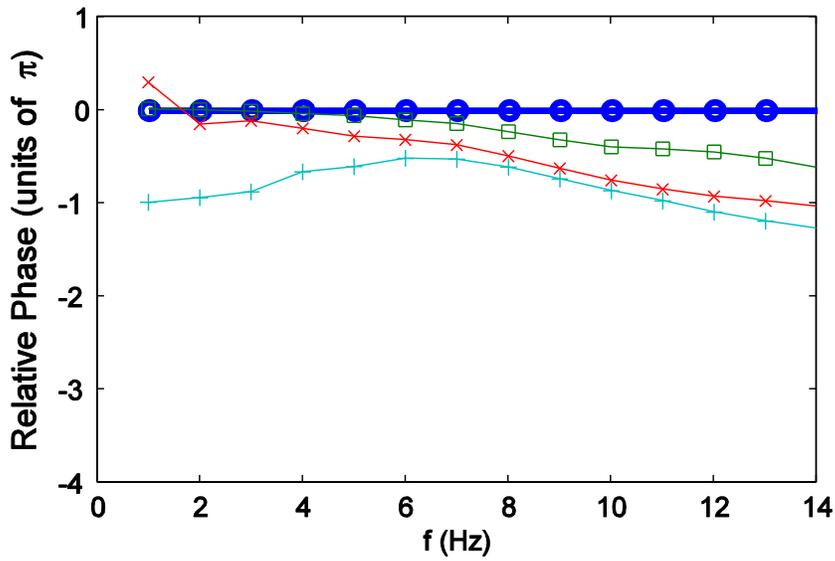

(c)

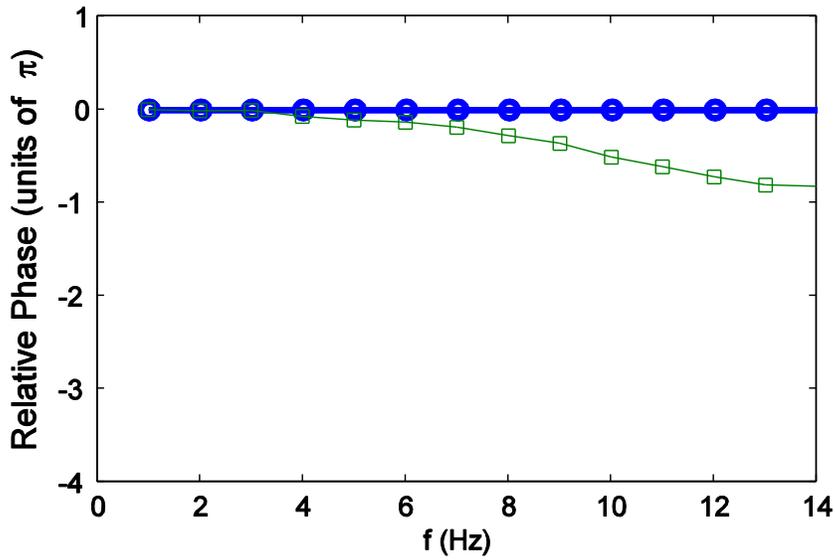

(d)

Figure 5. (Color online) Relative oscillation phase versus driving frequency. The phase is measured with respect to the phase of the particle in the string located farthest from the lower electrode. As shown, strings contain (a) eight particles, (b) six particles, (c) four particles,

and (d) two particles. Note that in (a) – (c) two distinct oscillation phases can be seen at low frequencies. In each case, P1 and P2 are π out of phase (at low frequencies) with the remainder of the particles in the string.

As shown, as the frequency approaches zero ($\omega \to 0$), the particles within each string separate into two distinct phase groups. The *zero-group*, consisting of P1 and P2, is out of phase by π rad with the *π-group*, which is comprised of the remaining particles in the string. The single exception to this appears for a four-particle string (Figure 5(c)), in which P3 joins the *zero-group* at low frequencies.

## III. ANALYSIS AND DISCUSSION

A number of interesting trends can be seen in the data above. As shown in Figure 4, across strings of different length the resonance peak amplitudes, frequencies, and damping coefficients remain consistent for a given particle position within different sized strings. Closer examination also shows that: particles P1 and P2 exhibit the highest damping coefficients, almost twice the magnitude of that found for P3 to P8; resonant frequencies show a linear decrease past the third particle; shorter strings (less than eight particles) behave as truncated segments of longer strings; and at low frequencies, dust particles separate into two phase groups having a phase difference of π rad. Each of these results is discussed below.

*Equilibrium positions and dispersion relations*

For small amplitude oscillations around the equilibrium positions, using a linear approximation and assuming only nearest neighbor interactions, the dust particles' equations of motion can be written as [22],

$$m_d \frac{d^2 \delta z_n}{dt^2} = -\gamma \delta z_n + \frac{Q^2}{4\pi\varepsilon_0 r_0^3} e^{-r_0/\lambda_D} \left(1 + r_0/\lambda_D\right)\left(2\delta z_n - \delta z_{n-1} - \delta z_{n+1}\right) \quad (2)$$

Here $z_n$ is the n[th] particle's equilibrium position, $\gamma = -\left(E\, dQ/dE + Q\, dE/dz\right)$, $\lambda_D$ is the screening length, and $r_0$ is the separation between particles. From Eq. 2 the dispersion relation can now be derived as:

$$\omega^2 = \frac{\gamma}{m_d} - \frac{4Q^2}{4\pi\varepsilon_0 m_d r_0^3} e^{-r_0/\lambda_D} \left(1 + r_0/\lambda_D\right) \sin^2 \frac{kr_0}{2}. \quad (3)$$

It is important to note that in Eq. 2 ignores the neutral drag term, $m_d \beta (\delta \dot{z}_n)$, which if included introduces a pure imaginary part to Eq. 3, $\omega = \omega_r + i\omega_i$ and $k = k_r + ik_i$. Since these imaginary parts contribution only to the overall transient behavior or instability, which are not the focus of this paper, only the real terms in the dispersion relations will be discussed here.

In order to properly examine each of the above, both the potential and the charge distribution in the sheath must first be considered in greater detail. A self-consistent fluid model was employed to determine the potential and plasma profiles along the central, vertical axis of the discharge chamber for the operating parameters used in the experiment. The model used solves the continuity equation,

$$\frac{\partial n_{e,i}}{\partial t} + \nabla \cdot \Gamma_{e,i} = S_{e,i} \quad (4)$$

for each plasma species including source terms. Rather than solving the momentum equation, the drift diffusion approximation is used to calculate the particle fluxes, $\Gamma_{e,i}$. The energy balance for electrons is also solved using a similar drift-diffusion approximation while the ion energy is assumed to be locally dissipated through frequent collisions with neutral atoms. A complete explanation of the fluid model can be found in [23, 24]. The resulting profiles for the electron density, ion density, electron temperature, and ion velocity were used in the current balance equation to obtain the dust potential as a function of height above the lower electrode, z, with the results shown in Fig 6(a). Finally, the dust potential was used assuming a capacitance model to calculate the dust charge number as a function of z. The calculated dust charge and potential distributions in the sheath are shown in Fig 6(b).

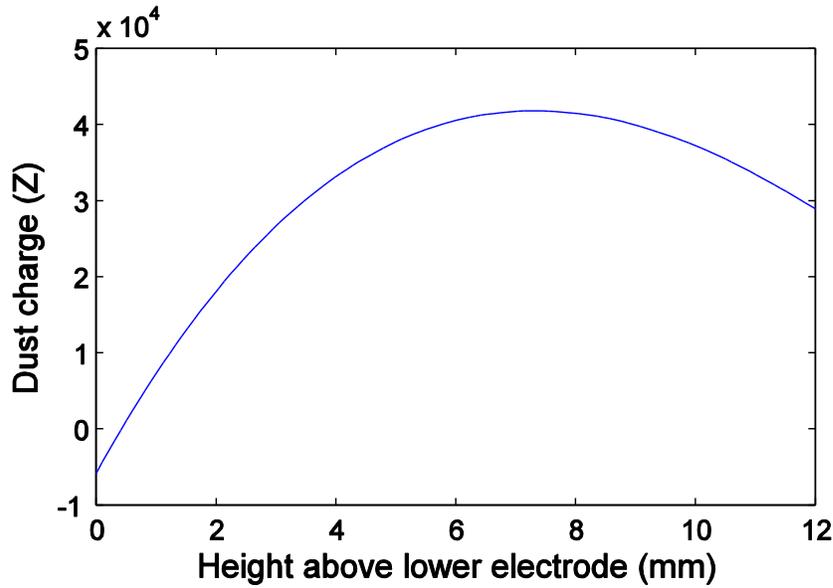

(a)

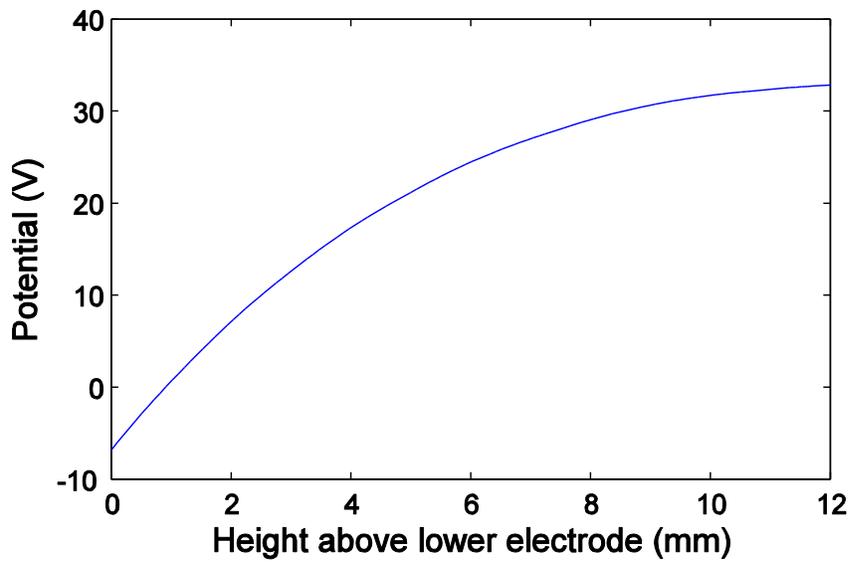

(b)

Fig 6. Calculated charge (a) and potential (b) distributions in the sheath.

*Resonance frequency*

Based on Eq 3 and the data shown in Fig 6, the resonance frequency distribution was calculated and then compared with experiment data (Fig 4(b)). The result is shown in Fig 7.

In Fig 7(a), the experimental data of phase difference at 9 Hz were fitted linearly and extrapolated to obtain the wavelength and thus the wave number $k$. For this case, the average particle separation was $r_0$ = 560 μm, with $\lambda_D$ directly derived from the potential simulation. These values were then inserted into Eq. 3 to obtain the frequency distribution as a function of height above the lower electrode (Fig 7(b)).

The experimental values shown exhibit a similar trend to the calculated frequency distribution. The small discrepancy between the experimental and the calculated data can be contributed to an ion drag effect [25 – 29]. A positive charge region created by streaming ions and formed behind the stationary dust particles can be described by an equivalent electric field $E_{ion}$, which reduces the total E field and lowers the oscillation frequency distributions.

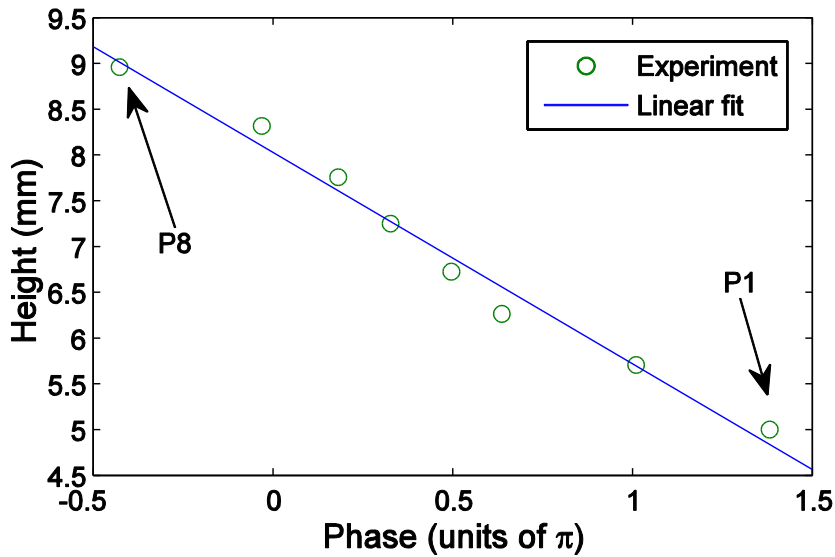

(a)

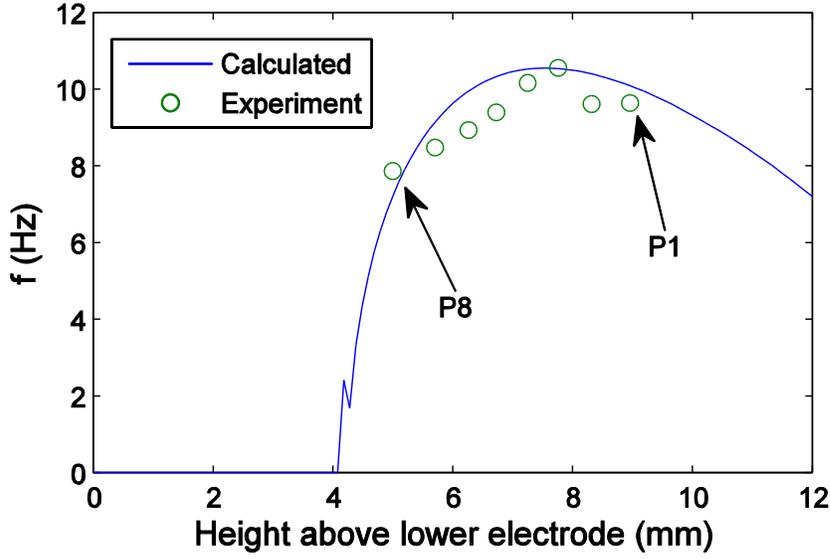

(b)

Fig 7. (Color online) (a) Extrapolated linear fit to obtain the wavelength. (b) Calculated frequency distribution compared with experimental data.

*Phase difference as ω → 0*

Finally as shown in Figure 5, at lower driving frequencies, particles within each string separate into two distinct phase groups with the two particles located farthest from the lower electrode, P1 and P2, out of phase by π rad with the rest of the particles in the string. By definition, the oscillation phase difference between any two particles is given by

$$\tan(\varphi_2 - \varphi_1) = \frac{\tan(\varphi_2) - \tan(\varphi_1)}{1 + \tan(\varphi_2)\tan(\varphi_1)} \quad (5)$$

where $\tan(\varphi_i) = \frac{\beta_i \omega}{\omega_{0i}^2 - \omega^2}$. As ω → 0, tan(φ$_i$) and tan(φ$_2$ − φ$_1$) → 0, and (φ$_2$ − φ$_1$) → nπ (n = 0, ±1, ±2, …). Therefore, mathematically both the zero- and π-phase groups are solutions to Eq. (4) for oscillation frequencies approaching zero. Since ω → 0 is the equilibrium condition, the equilibrium condition, $QE = m_d g$, must be expanded to include additional terms. The sheath electric field, the interaction Yukawa field, and the ion drag effect must all be included [25]. Therefore, $QE = Q(E_{sheath} + E_{int} + E_{ion}) = m_d g$, where $E_{ion}$ is the equivalent ion drag electric field. Assuming that a positive ion-rich region is created by the ion focusing effect caused by the streaming ions, and that this region is located between P2 and P3, the force exerted on the upper (P1 and P2) and lower (P3 – P8) particles will be in opposite directions. Such an effect would cause a difference in phase similar to that seen experimentally between the π- and zero-phase groups.

As a first attempt to confirm this theory, an external DC bias was applied to the lower electrode and then varied in an attempt to modify the proposed ion focus region. Preliminary results from this experiment are shown in Fig 8.

As shown, the two phase groups are directly impacted by varying the DC bias. Since this should also affect the shape and/or position of the proposed ion focusing region, one preliminary assumption that could be made is that the two phase groups seen experimentally are created by this effect. Additional research to determine whether this assumption is correct is underway.

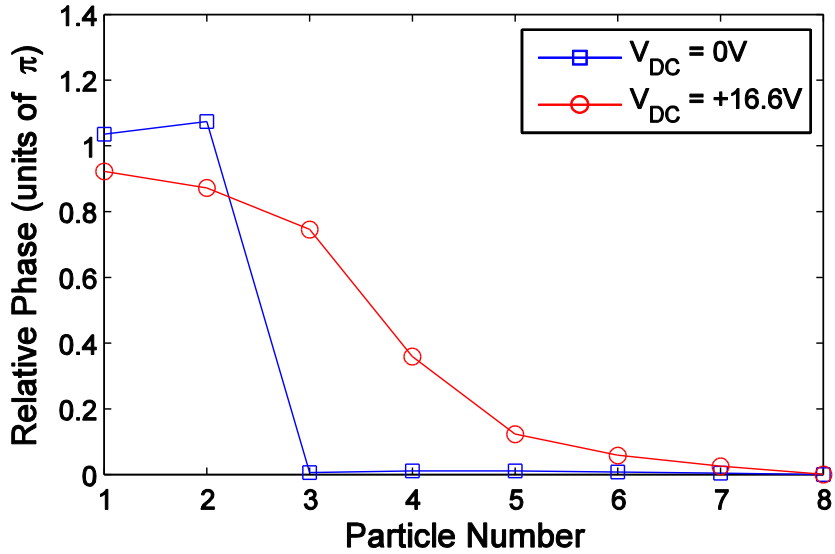

Figure 8. (Color online) Phase distributions for an eight-particle string under two DC biases applied externally to the lower electrode. In the case of an applied DC bias $V_{DC} = 0V$, two clearly separated phase groups can be seen. When the DC bias increased to $V_{DC} = +16.6V$ a gradual phase change occurs exhibiting a maximum slope w between P3 and P5. For both cases, the RF power is 2W, the natural DC bias is -17.8V, an the driving frequency is f = 1Hz.

*Damping*

Assuming Epstein drag, the damping coefficient $\beta$ will be proportional to the inverse of the dust particle radius, i.e. $\beta \propto 1/r_d$. As such, smaller dust particles should have higher $\beta$ values. Experimentally measured damping coefficients for a single layer of 8.89 μm dust particles levitated above the lower electrode (but with no glass box in the system) fall in the range $\beta \sim 12 - 20$ [30]. As shown in Figure 4(c), for the case at hand $\beta \sim 17$, in agreement with these measurements, except for P1 and P2. Possible causes for the higher damping coefficient exhibited by particles P1 and P2 include, a) the possibility that they have smaller particle radii (this would require they have half their measured radius, which is unlikely since the MF particles employed by this study are manufactured to a very high tolerance), or b) some other effect, for example, an ion rich region produced by the oscillation of the particles. The latter of these possibilities is currently under investigation.

IV. CONCLUSIONS

A mechanism for forming vertical, one-dimensional dust particle strings inside a glass box placed on the lower powered electrode of a plasma discharge device has been described. The dynamic properties of the vertical string have been studied by examining the oscillation spectra for one-dimensional strings of various lengths through modulation of the lower electrode DC bias. The resulting particle oscillation amplitudes, resonance frequencies, damping coefficients and oscillation phase differences were compared. The wave properties of these vertical strings have also been examined. Our results show that the vertical interaction forces between the dust particles are important for the overall dynamic properties of the vertical string oscillations, such as the resonance frequency distributions, while the ion drag force most strongly effects the system close to equilibrium, creating the anomalous behavior and phase difference exhibited by P1 and P2. Although this effect is not yet completely understood, it may be related to an ion focusing effect.